\documentclass[usegraphicx,usenatbib]{mn2e}

\newcommand{\aap}{A\&A}
\newcommand{\apj}{ApJ}
\newcommand{\apjl}{ApJ}

\newcommand{\mnras}{MNRAS}

\newcommand{\pre}{Phys. Rev.}

\title[Chemical mixing in SPH simulations]{Chemical mixing in smoothed particle hydrodynamics simulations}

\author[Greif et al.]{Thomas H. Greif$^{1,2,3}$\thanks{E-mail: tgreif@ita.uni-heidelberg.de}, Simon C. O. Glover$^{1}$, Volker Bromm$^{2}$ and Ralf S. Klessen$^{1}$\\$^{1}$ Institut f\"{u}r Theoretische Astrophysik, Albert-Ueberle Strasse 2, 69120 Heidelberg, Germany\\$^{2}$ Department of Astronomy, University of Texas, Austin, TX 78712, USA\\$^{3}$ Fellow of the International Max Planck Research School for Astronomy and Cosmic Physics at the University of Heidelberg}

\begin{document}

\maketitle
\topmargin-1cm

\begin{abstract}
We introduce a simple and efficient algorithm for diffusion in smoothed particle hydrodynamics (SPH) simulations and apply it to the problem of chemical mixing. Based on the concept of turbulent diffusion, we link the diffusivity of a pollutant to the local physical conditions and can thus resolve mixing in space and time. We apply our prescription to the evolution of an idealized supernova remnant and find that we can model the distribution of heavy elements without having to explicitly resolve hydrodynamic instabilities in the post-shock gas. Instead, the dispersal of the pollutant is implicitly modelled through its dependence on the local velocity dispersion. Our method can thus be used in any SPH simulation that investigates chemical mixing but lacks the necessary resolution on small scales. Potential applications include the enrichment of the interstellar medium in present-day galaxies, as well as the intergalactic medium at high redshifts.
\end{abstract}


\section{Introduction}
Understanding chemical enrichment and the dispersal of heavy elements in the wake of energetic supernovae (SNe) has become essential to a number of fields in astrophysics. The details of how enriched material mixes with ambient gas are not only relevant for the cooling and fragmentation properties of the interstellar medium (ISM), but also manifest themselves in the composition and dynamics of the resulting stars. As massive stars return metals to the ISM, mixing plays a key role in the overall matter cycle in galaxies. Tracing this loop back in time, one encounters the initial enrichment of the pristine, pure H/He gas at the end of the cosmic dark ages, when the very first stars, termed Population III (or Pop~III), ended their lives in violent SN explosions and expelled a significant fraction of their mass in metals \citep{heger03,iwamoto05,greif07,wa08b}. The resulting mixing in the early Universe has two aspects. The first concerns the enrichment inside the first galaxies, when the metals ejected by Pop~III stars recollapsed into more massive haloes that cooled by atomic hydrogen lines and vigorously mixed with pristine material in the presence of a highly turbulent medium \citep{wa07b,greif08a}. The second relates to the enrichment of the intergalactic medium (IGM) at redshifts $z>5$ \citep{mfr01}. The resulting metallicity distribution is a crucial ingredient in modeling the reionization history of the Universe \citep{ybh04,fl05} and in determining when cosmic star formation shifted from a predominantly high-mass, Pop~III mode, to a more normal Pop~II mode \citep{schneider02,mbh03,venkatesan06,tfs07}.

Unfortunately, an accurate treatment of mixing, while simultaneously simulating the larger-scale environment, is presently not feasible, as the turbulent motions responsible for mixing typically cascade down to very small scales. A frequently encountered approach to this problem is to assume that the products of stellar nucleosynthesis are distributed within a fixed volume \citep[e.g.][]{nop04,scannapieco05b,ksw07,tfs07}. Significantly better results can be achieved in grid-based simulations by relating the mass flux between cells to the mixing efficiency, even though it remains unclear how much of this mixing is numerical instead of physical \citep[e.g.][]{wa08b}. Such a direct approach is difficult to implement in SPH simulations, due to their Lagrangian nature, and instead chemical mixing has been modelled as a diffusion process \citep[e.g.][]{martinez-serrano08}. This is somewhat more accurate since the rms displacement of a fluid element in a homogeneously and isotropically driven turbulent velocity field can be described by the diffusion equation, with the diffusion coefficient set by the velocity dispersion and the typical shock travel distance \citep{kl03}.

Prescriptions with a constant diffusion coefficient have been applied to models of the chemical evolution of the Milky Way \citep{karlsson05,kg05}, galaxies in a cosmological context \citep{martinez-serrano08}, and the environment of the first galaxies \citep{kjb08}. In the present paper, we introduce an improved method that resolves chemical mixing in space and time based on the velocity dispersion within the SPH smoothing kernel. This yields results in agreement with more detailed investigations of mixing in the early phases of SN remnants, without having to explicitly resolve the hydrodynamic instabilities in the post-shock gas. In future work, we plan to use our method to investigate the long-term evolution of energetic SNe in a cosmological environment, particularly at high redshift when the Universe was enriched with the first metals. However, since our algorithm is quite generic, we hope that it will prove a valuable tool for any SPH simulation that attempts to follow the mixing of pollutants.

The structure of our work is as follows. In Section~2, we describe our model for diffusion and its numerical implementation in the cosmological simulation code {\sc GADGET}-2 \citep{springel05}, followed by a series of idealized test simulations (Section~3). We then discuss the relation between the diffusion coefficient and the velocity dispersion and apply our prescription to the evolution of an idealized SN remnant (Section~4). Finally, in Section~5 we summarize our results and assess their implications. For consistency, all quoted distances are physical, unless noted otherwise.

\section{Diffusion algorithm}
Diffusion plays an important role in a variety of astrophysical contexts. Among them are thermal conduction \citep[e.g.][]{jsd04}, heat transfer in shear flows \citep[e.g.][]{wvc08}, the microscopic diffusion of particles, such as photons in stellar interiors, or the spatial correlation of individual fluid elements in a turbulent medium \citep{kl03}. These processes are all described by the diffusion equation, commonly written in the form:
\begin{equation}
\frac{{\rm d}c}{{\rm d}t}=\frac{1}{\rho}\nabla\cdot(D\nabla\,c)\mbox{\ ,}
\end{equation}
where $c$ is the concentration of a contaminant fluid per unit mass, $D$ is the diffusion coefficient, which can be a function of space and time, and ${\rm d}/{\rm d}t$ the Lagrangian derivative, or the derivative following the motion. The diffusion coefficient has dimensions $\rm{M}\,\rm{L}^{-1}\,\rm{T}^{-1}$, suggesting that it can be represented as the product of the local density and some typical length-scale and velocity, such as the particle mean free path and velocity dispersion in a microscopic picture of diffusion.

\subsection{Numerical implementation}
In the SPH formalism, the diffusion equation can be reduced to a discrete summation over all particles within the smoothing kernel:
\begin{equation}
\frac{{\rm d}c_{i}}{{\rm d}t}=\sum_{j}K_{ij}(c_{i}-c_{j})\mbox{\ ,}
\end{equation}
with
\begin{equation}
K_{ij}=\frac{m_{j}}{\rho_{i}\rho_{j}}\frac{4D_{i}D_{j}}{(D_{i}+D_{j})}\frac{\textbf{r}_{ij}\cdot\nabla_{i}W_{ij}}{r_{ij}^{2}}\mbox{\ ,}
\end{equation}
where $i$ and $j$ denote the particle indices, $m$ the mass, $\rho$ the density, $W_{ij}$ the kernel and $\textbf{r}_{ij}$, $r_{ij}$ the vector and absolute separations between particles $i$ and $j$, respectively \citep[for a more detailed derivation, see][]{mhw05}. In the above equation, the arithmetic mean of the diffusion coefficient has been replaced by the harmonic mean, which has proven to be more robust. Furthermore, the second derivative has been replaced by a term involving the gradient and the particle separation, since a direct computation of the second derivative is problematic \citep{monaghan05}.

The solution to equation~(2) can either be determined explicitly, which requires an additional constraint on the time-step to ensure numerical stability, or implicitly, which requires the solution of a coupled set of differential equations involving all `active' particles (i.e. all SPH particles that are being updated on the current time-step). The explicit approach is extremely difficult to implement in a conservative fashion in an SPH code that allows individual particles to have different time-steps (e.g. {\sc GADGET}-2). This is because in a code of this type, neighbouring particles will sometimes have different time-steps. Consequently, the increase in concentration at particle $i$ caused by diffusion from particle $j$ will sometimes be computed at a different time from the corresponding decrease in concentration at particle $j$ caused by diffusion to particle $i$. To ensure conservation, the increase and decrease must exactly balance, but in general they will not if $i$ and $j$ have different time-steps. One could, of course, avoid this problem by ensuring that all particles are synchronized before their concentrations are updated. However, the required synchronizations would have to occur very frequently, and so one would lose essentially all of the benefits gained by allowing the particles to have individual time-steps. A further undesirable feature of the explicit approach is the fact that the time-steps required to stably model the diffusion can become very small. Consideration of equation~(1) shows that the required time-step scales with the spatial resolution -- represented in an SPH code by the smoothing length $h$ -- as
\begin{equation}
\Delta t \propto h^{2}\mbox{\ .}
\end{equation}
In comparison, the standard Courant time-step scales only linearly with $h$.

An implicit approach to the solution of equation~(2) avoids some of these problems, as it allows one to take larger time-steps without compromising numerical stability. However, this comes at a cost: the coupled differential equations representing the diffusion must be solved iteratively, and it is difficult to do this in a fashion that can be efficiently parallelized. In addition, one still has to deal with the synchronization problem discussed above.

In view of the disadvantages of both standard approaches, it is interesting to explore the viability of simpler, but more approximate approaches, such as the one presented in this paper. To obtain our approximation, we assume that the densities and concentrations of all active particles do not change significantly over a time interval $\Delta t$, allowing a direct integration of equation~(2):
\begin{equation}
c_{i}(t_{0}+\Delta t)=c_{i}(t_{0})e^{A\Delta t}+\frac{B}{A}(1-e^{A\Delta t})\mbox{\ ,}
\end{equation}
with
\begin{equation}
A=\sum_{j}K_{ij}
\end{equation}
and
\begin{equation}
B=\sum_{j}K_{ij}c_{j}\mbox{\ .}
\end{equation}
For large time-steps, the concentration of particle $i$ thus tends to the average among its neighbours, while for small time-steps it remains close to its original concentration.

We implement this approach by performing the required operations at a global synchronization point in the density routine of {\sc GADGET}-2. After the new densities and smoothing lengths have been computed, we update the coefficients $K_{ij}$ and subsequently use equation~(5) to determine the new concentrations of all active SPH particles. In a final step, we renormalize the obtained concentrations with a global factor such that the total concentration is conserved. Since the Courant condition does not allow for significant changes in density between time-steps, and diffusion generally progresses slower than the speed of sound, our implementation is quite generic and can be applied to a number of problems in astrophysics. We have found that the algorithm is remarkably stable even for very short diffusion timescales, since particles tend to equilibrate their concentrations and neighbouring particles are generally active at the same time. The additional CPU consumption is minimal since we utilize the pre-existing neighbour search. By the same token, the algorithm is easy to implement in any SPH code and is not restricted to {\sc GADGET}-2.

\subsection{Test problems}
In this section we investigate the formal accuracy of the algorithm by performing a number of idealized test simulations. We initialize all simulations in a periodic, uniform density box with $1$ million SPH particles and length and sound-crossing time set to unity, such that we may conveniently quote the elapsed time in units of the sound-crossing time. We adopt a hydrogen number density of $n_{\rm{H}}=1~\rm{cm}^{-3}$ and a mean molecular weight corresponding to that of a neutral, primordial gas. We place the particles on a grid with a very small random displacement, and in the first test problem also consider a fully random distribution of particles, such that the density fluctuates considerably around the mean. This gives a better handle on the performance of our algorithm under more realistic circumstances. In all cases, we use the maximum diffusivity that is accurately modelled by our algorithm, determined by the Courant condition (see Sections 2.1 and 3.1):
\begin{equation}
D=\rho\,h\,c_{\rm{s}}\mbox{\ ,}
\end{equation}
where $c_{\rm{s}}$ is the sound speed and $\rho$ and $h$ are determined by the mean density $n_{\rm{H}}=1~\rm{cm}^{-3}$, such that the diffusion coefficient becomes a fixed numerical value.

In the first test problem, we consider the propagation of an initial $\delta$-distribution, to which the analytic solution is Green's function of the diffusion equation:
\begin{equation}
G(\textbf{x},\textbf{x}',t)=\frac{1}{(2\pi\sigma^{2})^{3/2}}\exp\left({{{\frac{-|\textbf{x}-\textbf{x}'|^{2}}{2\sigma^{2}}}}}\right)
\end{equation}
with variance
\begin{equation}
\sigma = \sqrt{2Dt}\mbox{\ .}
\end{equation}
This configuration is reproduced by setting the concentration of the central particle to unity, and of all others to zero. In the left-hand column of Fig.~1, we compare the analytic solution to the simulation results at three different output times. Early on, diffusion progresses somewhat too rapidly, since the concentrations of neighbouring particles differ substantially. In this case, equation~(5) effectively breaks down, but the resulting deviations remain small and do not influence the long-term behavior, where all three curves become nearly indistinguishable. Note that the random distribution performs almost as well as the grid-based distribution, showing that the diffusivity remains unchanged even for high density fluctuations.

The second test problem consists of two individual $\delta$-distributions initially separated by $\Delta x=1/3$. The analytic solution can be obtained by a convolution of Green's function with the initial state of the system:
\begin{equation}
c(\textbf{x},t_{0}+\Delta t)=\int G(\textbf{x},\textbf{x}',t_{0}+\Delta t)\,c(\textbf{x}',t_{0})\,d\textbf{x}'\mbox{\ ,}
\end{equation}
such that the general solution is given by the superposition of two Gaussian distributions centred at $x=1/3$ and $2/3$. Similarly, the solution to the third test problem, consisting of a slab of uniform concentration between $x=1/3$ and $2/3$, is given by:
\begin{equation}
c(\textbf{x},t)=\frac{1}{2}\,{\rm erfc}\left(\frac{1/3-x}{\sqrt{4Dt}}\right)
\end{equation}
for $x\le 1/2$ and
\begin{equation}
c(\textbf{x},t)=\frac{1}{2}\,{\rm erfc}\left(\frac{x-2/3}{\sqrt{4Dt}}\right)
\end{equation}
for $x>1/2$. In the middle and right-hand columns of Fig.~1, we compare the simulation results to the analytic solution, showing that there are only minor deviations at early times, similar in nature to those found in the first test problem. A slice through all three boxes (Fig.~2) shows the solution in two dimensions.

Finally, in a series of resolution studies performed with $32^{3}$, $64^{3}$ and $128^{3}$ particles, we have found no correlation between the resolution and the deviation from the analytic solution. In fact, in all cases, the deviations were comparable to those found in previous test problems. Considering its formal simplicity, the algorithm thus performs remarkably well. Although the diffusivity is initially slightly over-predicted for cases in which the diffusion time is shorter than the sound-crossing time, we nevertheless find a correct long-term behavior.

\begin{figure*}
\begin{center}
\resizebox{17cm}{16cm}
{\includegraphics{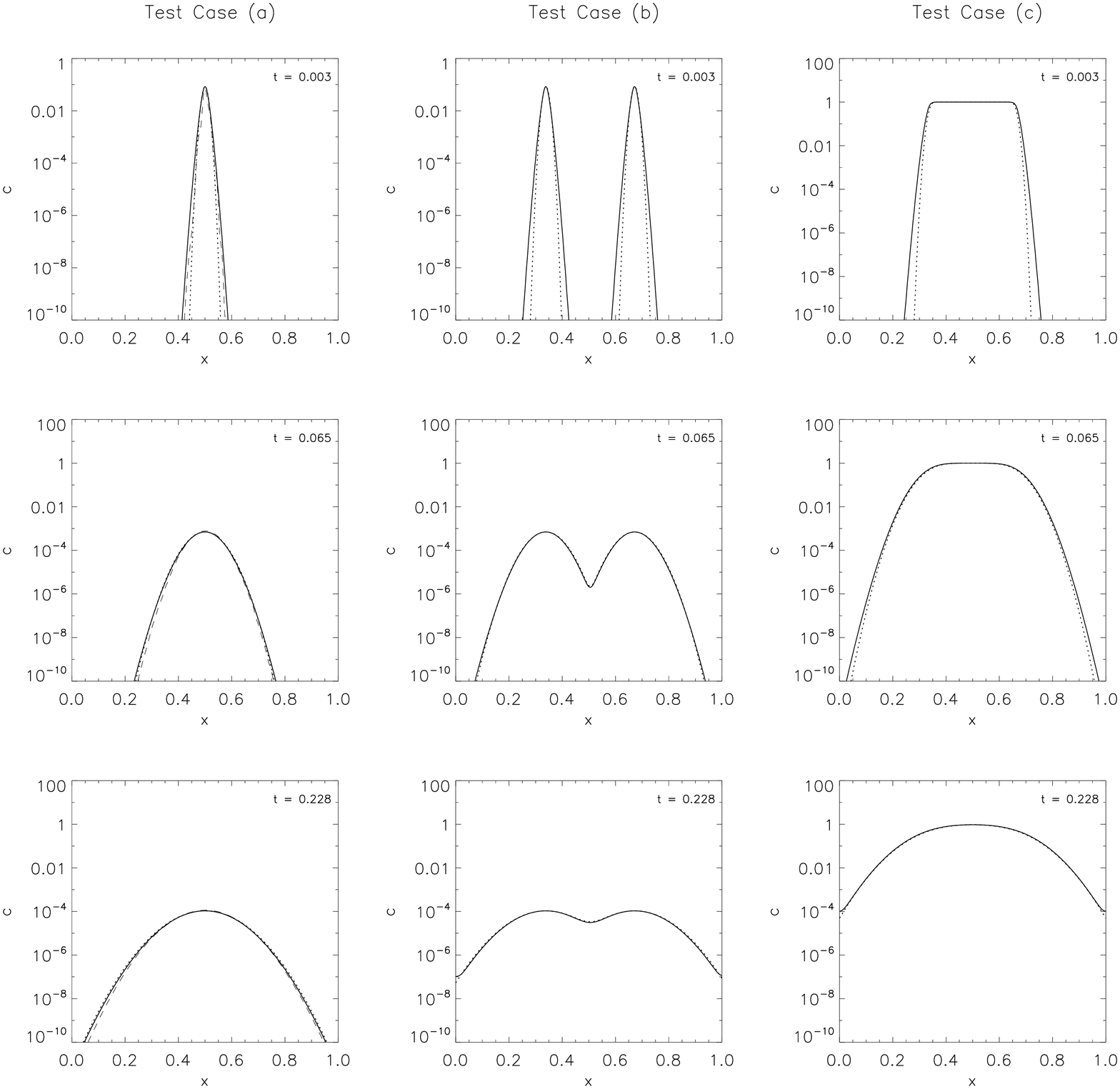}}
\caption{The propagation of a single $\delta$-distribution (left-hand column), two $\delta$-distributions (middle column) and a slab of uniform concentration (right-hand column) into an otherwise pristine medium, shown for a grid-based particle distribution (solid lines) and a random particle distribution (dashed line), compared to the analytic solution (dotted lines). The temporal evolution is depicted from top to bottom and quoted in units of the sound-crossing time. In all three cases, the simulations reproduce the analytic solution, except at early times when the underlying assumption of constant concentration for neighbouring particles is not well justified (see Section~2). However, as is evident from the bottom two rows, this does not affect the long-term behavior.}
\end{center}
\end{figure*}

\begin{figure*}
\begin{center}
\resizebox{17cm}{7.5cm}
{\includegraphics{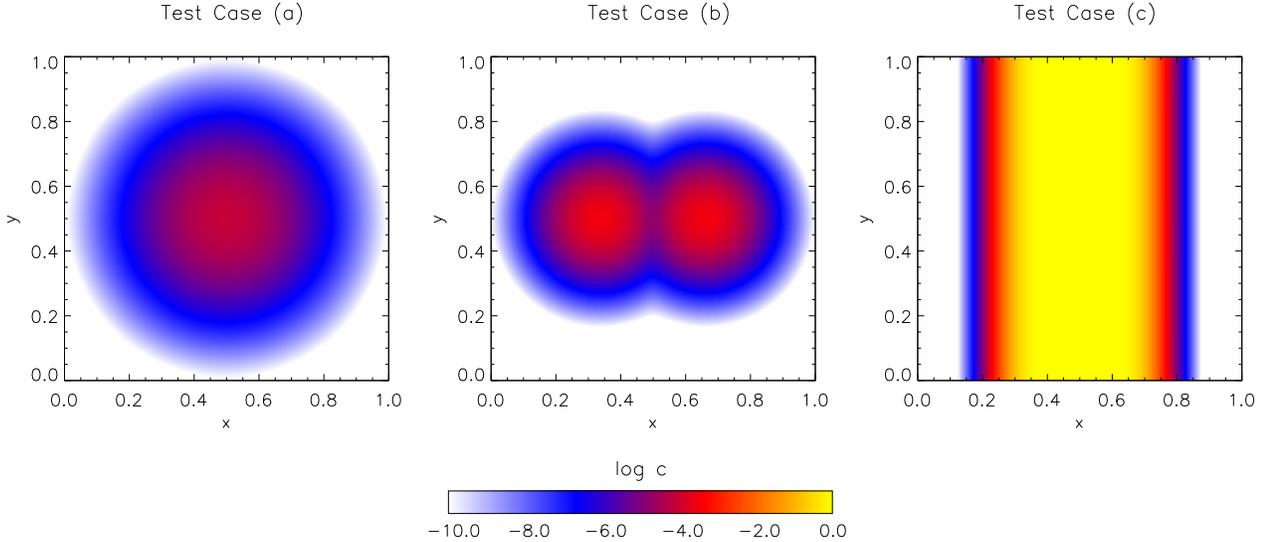}}
\caption{The propagation of a single $\delta$-distribution (left-hand column), two $\delta$-distributions (middle column) and a slab of uniform concentration (right-hand column) into an otherwise pristine medium, shown at a representative output time.}
\end{center}
\end{figure*}

\begin{figure*}
\begin{center}
\resizebox{17cm}{16cm}
{\includegraphics{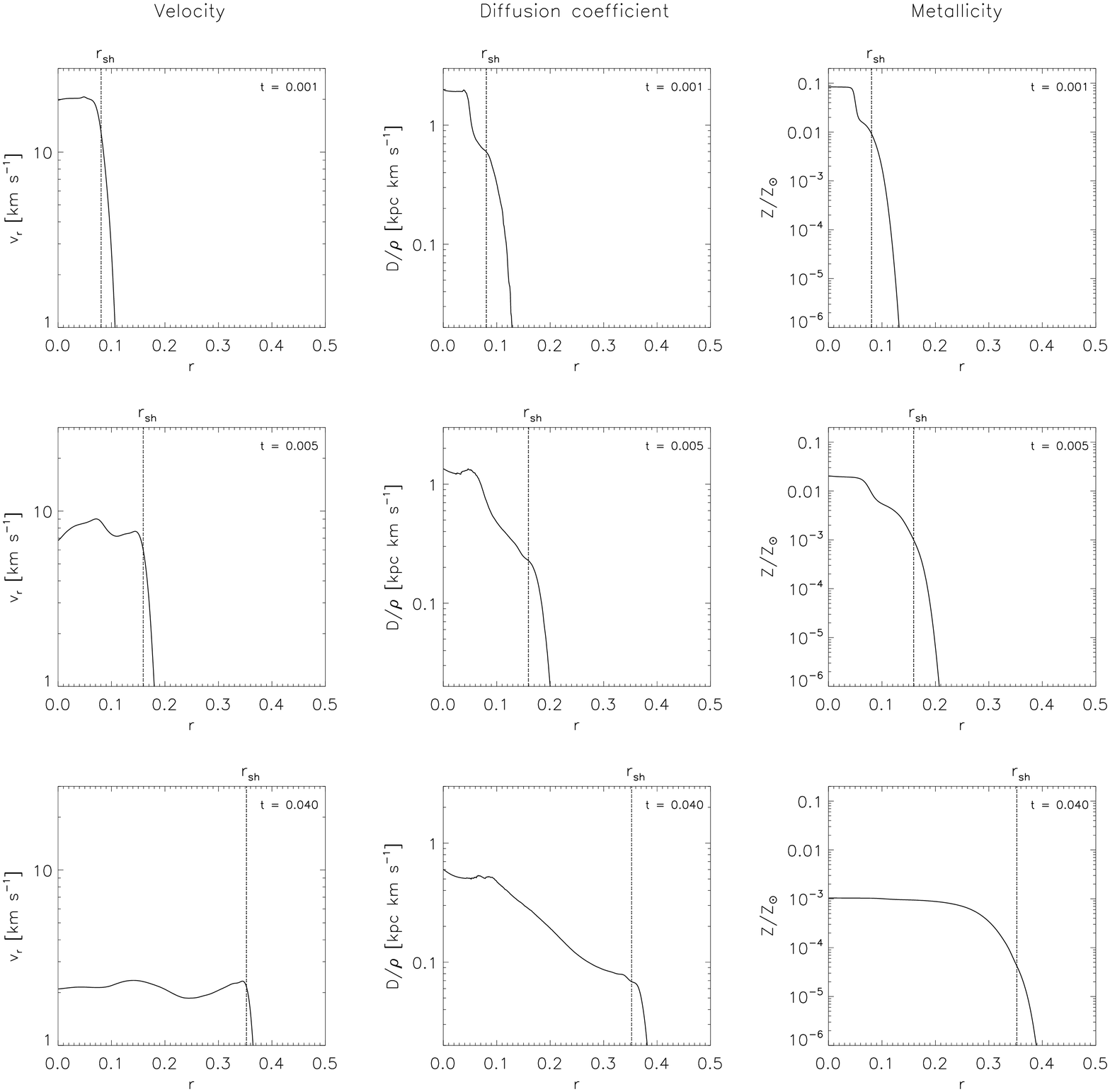}}
\caption{The radial velocity (left-hand column), diffusion coefficient (middle column) and metallicity (right-hand column) as a function of distance from the SN progenitor, with the position of the forward shock denoted by the dotted line. The temporal evolution is depicted from top to bottom and quoted in units of the sound-crossing time. The high velocity dispersion near the main shock leads to an elevated diffusion coefficient and thus very efficient chemical mixing. Due to the dependence on velocity dispersion, mixing by hydrodynamic instabilities in the remnant is implicitly accounted for.}
\end{center}
\end{figure*}

\section{Application to chemical mixing}
In the previous section, we introduced an SPH formalism for diffusion that can be applied to most problems that are governed by a diffusion equation. In this section, we focus on an application that is particularly relevant to astrophysics: the mixing of chemical elements.

\subsection{Chemical mixing as turbulent diffusion}
As a first step towards a model for chemical mixing, one must find a connection between the diffusivity of a pollutant and the local physical conditions. \citet{kl03} have provided this link by showing that the probability of finding a parcel of gas at a given location in a homogeneously and isotropically driven turbulent velocity field can be described by the diffusion equation, with the diffusion coefficient set by the velocity dispersion $\tilde{v}$ and the turbulent driving length $\tilde{l}$:
\begin{equation}
D=2\,\rho\,\tilde{v}\,\tilde{l}\mbox{\ .}
\end{equation}
This corresponds to the classical mixing-length approach extended into the supersonic regime. If one replaces the probability distribution with the concentration of a pollutant, this formalism can be reinterpreted to describe chemical mixing. The sole remaining task is then to provide the parameters $\tilde{v}$ and $\tilde{l}$ to the diffusion algorithm, such that the diffusion coefficient can become a function of space and time.

To obtain the necessary parameters, we determine the rms velocity dispersion for particle $i$ within its smoothing length:
\begin{equation}
\tilde{v}_{i}^{2}=\frac{1}{N_{\rm{ngb}}}\sum_{j}|\textbf{v}_{i}-\textbf{v}_{j}|^{2}\mbox{\ ,}
\end{equation}
where $N_{\rm{ngb}}$ is the number of neighbours and $\textbf{v}_{i}$ and $\textbf{v}_{j}$ are the bulk velocities of particles $i$ and $j$. Finally, we equate the turbulent driving length with the smoothing length, since this is the minimum scale where turbulent motions can be resolved. This then yields for the diffusion coefficient:
\begin{equation}
D_{i}=2\,\rho_{i}\,\tilde{v}_{i}\,h_{i}\mbox{\ .}
\end{equation}
As desired, the efficiency of chemical mixing is now governed entirely by local physical quantities.

An important underlying assumption of this method is that the velocity field on subresolution scales corresponds to a homogeneously and isotropically driven turbulent medium, i.e. only the magnitude of the velocity dispersion and not its three-dimensional structure is taken into account. A further implicit assumption is that the mixing efficiency on subresolution scales is set by the corresponding value on the scale of the smoothing kernel. For these reasons our method yields an upper limit to the amount of mixing that can occur, and does not capture the details of a real turbulent cascade. However, since turbulent motions on scales larger than the smoothing length are explicitly resolved, our approach should suffice for most practical purposes.

We implement the above steps in our algorithm by performing a previous neighbour search that finds the velocity dispersion for all active SPH particles, which is then used in equation~(3) to obtain the coefficients $K_{ij}$. This closes the required set of equations, and in the following subsection we use our complete model to investigate the mixing of metals in a SN remnant.

\subsection{Mixing in supernova remnants}
The mixing of gas in the post-shock regions of SN remnants has previously been investigated, leading to the consensus view that secondary shocks trigger Rayleigh-Taylor and Kelvin-Helmholtz instabilities that mix pristine gas from the dense shell with already enriched material \citep[e.g.][]{gull73,cbe92}. The aim of our algorithm is to capture this mixing without having to explicity resolve the corresponding hydrodynamic instabilities, which cannot easily be modelled using SPH \citep[e.g.][]{agertz07,price08}. To verify its ability to do this, we perform a test simulation of an idealized SN explosion with the same setup as in Section~2.2. We distribute an explosion energy of $10^{51}~\rm{erg}$ as thermal energy to the $N_{\rm{ngb}}$ nearest neighbours around the centre of the box, which results in the formation of a shock and reproduces the ideal Sedov-Taylor solution of a blast wave in a uniform medium \citep[see][]{greif07}. Furthermore, we set the initial metallicities of the central particles to unity.

In Fig.~3, we show the radial velocity, diffusion coefficient and metallicity as a function of distance from the SN progenitor at three different output times. The high velocity dispersion near the forward shock is responsible for the elevated diffusion coefficient, which leads to efficient chemical mixing of pristine gas passing through the shock. As is evident from Fig.~3, most parcels of gas behind the shock have nearly equilibrated their metallicities. Our algorithm thus implicitly accounts for mixing due to the dependence of the diffusion coefficient on the local velocity dispersion. Intriguingly, we also find metals ahead of the forward shock. Although this initially seems unphysical, it should be remembered that our algorithm aims to capture mixing caused by unresolved instabilities, and that in a real SN remnant, it is these instabilities that would transport metals ahead of the mean position of the shock. Even further out, the degree of enrichment drops roughly exponentially, since the velocity dispersion, and hence the diffusion coefficient, tend to zero.

A second mechanism becomes important once the SN remnant has stalled: gas from the dense shell ahead of the shock falls back on to the remnant, becomes Rayleigh-Taylor unstable, and vigorously mixes with the interior gas. This is especially effective in a realistic cosmological setting, where the central potential well deepens as the SN remnant expands \citep{greif07,wa08b}. However, we postpone a detailed treatment of this issue to dedicated high-resolution simulations in future work.

\section{Summary and conclusions}
We have introduced a simple and efficient algorithm for diffusion in SPH simulations and investigated its accuracy with a number of idealized test cases. Although the algorithm is quite generic and can be applied to most problems involving diffusion, we have here focused on a model for the dispersal of enriched material in supernova explosions. Adopting the mixing length approach discussed by \citet{kl03}, we link the diffusivity of the pollutant to the local physical conditions and can thus describe the space- and time-dependent mixing process. We have applied our prescription to an idealized SN explosion and found that we can properly describe the mixing process without having to explicitly resolve the hydrodynamic instabilities in the post-shock gas. Instead, this process is implicitly governed by the dependence on the local velocity dispersion. Our method can thus be used in any SPH simulation that attempts to follow the mixing of a pollutant but lacks the necessary resolution on small scales. This will be relevant, in particular, for simulations that aim at simultaneously representing the large-scale environment of the SN explosion and the fine-grained mixing. A crucial assumption of our model is that resolved motions cascade down to subresolution scales which then homogeneously mix the gas. For this reason our method yields an upper limit to the mixing efficiency and the resulting degree of homogeneity, and cannot capture the details of a real turbulent cascade. However, since turbulent motions on scales larger than the smoothing length are explicitly resolved, this approximation should generally be valid.

Our new algorithm ties in well with the current generation of radiation-hydrodynamical simulations of star formation, both at high redshifts and in the local Universe. The ultimate goal is to describe the interrelated cycle of gas fragmentation and stellar feedback. To achieve adequate realism, radiative effects have to be considered together with the mechanical and chemical feedback due to SN explosions. The methodology developed here allows us to include chemical feedback, suitable even for extremely large simulations. We will report on specific applications in future studies.

\section*{Acknowledgments}
The authors would like to thank the referee Joe Monaghan for his insightful comments and suggestions that greatly improved the quality of this work. TG is grateful to Paul Clark for many stimulating discussions on SPH. TG acknowledges financial support by the Heidelberg Graduate School of Fundamental Physics (HGSFP). The HGSFP is funded by the excellence initiative of the German government (grant number GSC 129/1). VB acknowledges support from NSF grant AST-0708795 and NASA grant NNX08AL43G. RSK thanks for partial support from the Emmy Noether grant KL 1358/1. RSK and TG also acknowledge subsidies from the DFG SFB 439, Galaxies in the early Universe. The simulations presented here were carried out at the Texas Advanced Computing Center (TACC).

\end{document}